\documentclass[twocolumn]{jpsj3} 
\usepackage{graphicx}
\newcommand{\ii}{{\rm i}}
\newcommand{\simg}{\stackrel{>}{_\sim}}
\newcommand{\siml}{\stackrel{<}{_\sim}}

\def\runauthor{Keisuke \textsc{Mitsumoto} and  Yoshiaki {\sc \=Ono}}

\title{Cooperative Effect of Coulomb Interaction and Electron-Phonon Coupling on the Heavy Fermion State in the Two-Orbital Periodic Anderson Model}

\author{Keisuke \textsc{Mitsumoto}$^{1}$\thanks{E-mail address: mitsumoto@phys.sc.niigata-u.ac.jp} and 
Yoshiaki \textsc{\=Ono}$^{2,3}$}

\inst{
$^{1}$ Graduate School of Science and Technology, Niigata University, Niigata 950-2181, Japan \\
$^{2}$ Department of Physics, Niigata University, Niigata 950-2181, Japan \\
$^{3}$ Center for Transdisciplinary Research, Niigata University, Niigata 950-2181, Japan}

\abst{
We investigate the two-orbital periodic Anderson model, where the local orbital fluctuations of $f$-electrons couple with a two-fold degenerate Jahn-Teller phonon, by using the dynamical mean-field theory. 
It is found that the heavy fermion state caused by the Coulomb interaction between $f$-electrons $U$ is largely enhanced due to the electron-phonon coupling $g$, in contrast to the case with the single-orbital periodic Anderson model where the effects of $U$ and $g$ compete to each other. 
In the heavy fermion state for large $U$ and $g$, both the orbital and lattice fluctuations are enhanced, while the charge (valence) and spin fluctuations are suppressed. 
In the strong coupling regime, a sharp soft phonon mode with a large spectral weight is observed for small $U$, while a broad soft phonon mode with a small spectral weight is observed for large $U$. 
The cooperative effect of $U$ and $g$ for half-filling with two $f$-electrons per atom $n_f=2$ is more pronounced than that for quarter-filling with $n_f=1$. 
}

\kword{
heavy fermion, 
periodic Anderson model, 
orbital degeneracy, 
Coulomb interaction, 
electron-phonon coupling, 
Jahn-Teller phonon, 
dynamical mean-field theory
}

\begin{document}
\maketitle
\newpage

\section{Introduction}

Heavy fermion systems have been intensively investigated for the last three decades. 
The effective mass of quasiparticles for the systems, estimated from the $T$-linear coefficient of specific heat, the Pauli paramagnetic susceptibility and the $T^2$ coefficient of resistivity, is much larger than that for simple metals. 
The strong correlation effects between $f$-electrons are considered to play crucial roles, where the large entropy due to the local magnetic moment of the $f$-electrons remains down to a low temperature $T_0$ and results in a large effective mass in proportion to $1/T_0$ at low temperature below $T_0$. 
When the degeneracy is resolved by external magnetic fields, the effective mass is expected to be suppressed. 
Such a magnetic suppression of the effective mass is widely observed in heavy fermion compounds, justifying the heavy fermion state of magnetic origin.

Recently, another class of heavy fermion systems has been observed in filled skutterudite compounds RT$_4$X$_{12}$ with R=rare earth, T=Fe, Ru, Os, and X=pnictogen, where the R ion is surrounded by an oversized cage made of X and T atoms. 
SmOs$_4$Sb$_{12}$ shows a large specific heat coefficient $\gamma =820\mathrm{mJ/K^{2}mol}$ which is almost independent of applied magnetic fields \cite{sanada}, suggesting a heavy fermion state of nonmagnetic origin such as charge, valence and phonon degrees of freedom. 
Sm ion in SmOs$_4$Sb$_{12}$ shows mixed valence state of Sm$^{3+}$ and Sm$^{2+}$ as observed in X-ray spectroscopy experiments \cite{mizumaki, yamasaki}. 
In the Sb-NQR experiment, a peak in $1/T_{2}$ has been attributed to some electrical fluctuations, probably local charge fluctuations \cite{kotegawa}. 
The specific heat and the ultrasonic measurements have revealed that the crystalline-electric-field (CEF) ground state of Sm$^{3+}$ ion is  $\Gamma_{67}$ quartet in T$_{h}$ site symmetry\cite{sanada,nakanishi}. 
However, another specific heat measurement claimed the $\Gamma_{5}$ doublet CEF ground state and there is still controversy about the CEF state\cite{yuhasz}. 

$\mathrm{PrOs_{4}Sb_{12}}$ with $\Gamma_{1}$ singlet ground state also shows a large specific heat coefficient $\gamma =750\mathrm{mJ/K^{2}mol}$ together with a large jump in the specific heat $\Delta C/T_{\mathrm{c}}\simg 500\mathrm{mJ/K^{2}mol}$ at the superconducting transition temperature $T_{\mathrm{c}}=1.85\mathrm{K} $\cite{bauer}. 
In the ultrasonic measurements, remarkable frequency dependence of the elastic constant (ultrasonic dispersion) around 30K has been observed in $\mathrm{PrOs_{4}Sb_{12}}$ and has been attributed to large amplitude local vibrations (rattling) of the Pr ion in the cage\cite{goto1}. 
In addition, $\mathrm{PrOs_{4}Sb_{12}}$ shows an anomalous softening of the elastic constant below 10K down to $T_{\mathrm{c}}$ \cite{goto1}. 
The softening is well accounted for by quadrupolar susceptibility due to $f$-electrons in the $\Gamma_1-\Gamma_4^{(2)}$ CEF state except for an extra softening below 3K \cite{goto2}, where the coupling between the quadrupolar fluctuations and the rattling may play important roles for the extra softening and also for the heavy fermion behavior down to $T_{\mathrm{c}}$. 
Similar properties of the ultrasonic dispersion and/or the anomalous softening have been observed also in other filled skutterudite compounds such as ROs$_4$Sb$_{12}$ (R=La, Sm, Nd) and RFe$_4$Sb$_{12}$ (R=La, Pr)\cite{yasumoto, nakanishi, yanagisawa, ishii1,ishii2}.

After the pioneering works of Holstein\cite{holstein}, the problem of coupling between local lattice vibrations (Einstein phonons) and local charge fluctuations of conduction electrons has been investigated by many authors. 
Yu and Anderson originally predicted that the strong electron-phonon coupling causes an effective double-well potential for oscillating ions which are responsible for the anomalously large resistivity and Debye-Waller factor observed in A15 compounds such as $\mathrm{V_{3}Si}$ and $\mathrm{Nb_{3}Ge}$\cite{anderson1,anderson2}. 
As a strong coupling fixed point, the two-level Kondo systems were investigated to describe a heavy-fermion like behavior observed in such compounds\cite{matsumi1,matsumi2}. 
Recently, the precise studies for the electron-phonon impurity models\cite{kusumi,yotsuhashi,hotta1,hewson1} and the multi-level impurity models\cite{hattori11, hattori12} have been done by using the numerical renormalization group (NRG) method and have confirmed the previous results including the double-well potential formation in the strong coupling regime. The similar approaches have also applied to solve the lattice models such as the Holstein model\cite{meyer} and the Hubbard-Holstein model\cite{koller} in infinite dimensions on the basis of the dynamical mean-field theory\cite{georges} (DMFT) and have revealed that the system becomes bipolaronic insulator when the electron-phonon coupling is larger than a critical value, $g> g_{c}$, where the effective potential becomes double-well type. 
More recently, Hattori and Miyake have discussed the ultrasonic dispersion in the Holstein model by using the self-consistent ladder approximation\cite{hattori2}.

To discuss the effect of local lattice vibrations on the rare-earth and actinide compounds, the periodic Anderson-Holstein model, where local phonons couple with valence fluctuations of $f$-electrons which interact with each other via the Coulomb interaction and hybridize with conduction electrons, was extensively studied in early theories for mixed-valence systems such as Sm chalcogenides\cite{varma}. 
As the strong correlation effect due to both the Coulomb interaction and the electron-phonon coupling is crucial for describing the heavy-fermion state in these systems, we need reliable and nonperturbative approaches such as the DMFT. 
Recently, we have studied the periodic Anderson-Holstein model by using the DMFT\cite{mitsumoto11, mitsumoto12, mitsumoto13}. 
What we found are as follows: 
(1) In the strong electron-phonon coupling regime, $g\simg g_{c}$, the system shows an anomalous heavy fermion behavior which is accompanied by a large lattice fluctuation and an extreme phonon softening. 
(2) A simple harmonic potential for ions for $g\siml g_{c}$ changes into an effective double-well potential for $g\simg g_{c}$. 
(3) The pairing interaction between the conduction electrons has a maximum at $g\approx g_{c}$.
(4) The heavy fermion state due to the electron-phonon coupling is realized in the wide range of the f-electron number $n_{f}$, while that due to the Coulomb interaction is realized in the narrow range near the half-filling $n_{f}\sim 1$. 
(5) The effect of the electron-phonon coupling on the heavy fermion state and that of the Coulomb interaction are compete with each other. 

In the rare-earth and actinide compounds, orbital fluctuations of $f$-electrons are considered to play crucial roles in determining the heavy-fermion behavior\cite{sato}. 
Recently, Hotta studied the multiorbital impurity Anderson model in which the orbital fluctuations of $f$-electrons couple with Jahn-Teller (JT) phonons by using the NRG method and observed the remarkable quasi-Kondo behavior due to the dynamical JT effect\cite{hotta2}. 
As for the lattice model, we have studied the two-orbital periodic Anderson model coupled with JT phonons by using the DMFT and have found that the heavy fermion state is realized due to the cooperative effect of the Coulomb interaction and the electron-phonon coupling\cite{mitsumoto2}. This is a striking contrast to the case with the periodic Anderson-Holstein model where the both effects are compete with each other as mentioned above \cite{mitsumoto11, mitsumoto12, mitsumoto13}. 
However, the effects of the degeneracy of JT phonon mode and the $f$-electron filling have not been discussed there.

In the present paper, we investigate the two-orbital periodic Anderson model, where JT phonons with a two-fold degenerate mode couple with orbital fluctuations of $f$-electrons which interact with each other via the Coulomb interaction and hybridize with conduction electrons, by using the DMFT to elucidate the cooperative effect of the Coulomb interaction and the electron-phonon coupling on the heavy-fermion state in the cases with half-filling with two $f$-electrons per atom $n_f=2$ and quarter-filling with $n_f=1$.

\section{Formulation}
Our model Hamiltonian is given by,
\begin{eqnarray}
H&=&\sum_{{\bf k}l\sigma}\epsilon_{{\bf k}}c^{\dag}_{{\bf k}l\sigma}c_{{\bf k}l\sigma}
  +\epsilon_{f}\sum_{il\sigma}f^{\dag}_{il\sigma}f_{il\sigma}  \nonumber \\
&+&V\sum_{il\sigma} (f^{\dag}_{il\sigma}c_{il\sigma}+h.c. ) \nonumber \\
&+&U\sum_{il}\hat{n}_{fil\uparrow}\hat{n}_{fil\downarrow} 
  +U'\sum_{i\sigma\sigma'}\hat{n}_{fi1\sigma}\hat{n}_{fi2\sigma'} \nonumber \\
&+&J\sum_{i\sigma\sigma'}f^{\dagger}_{i1\sigma}f^{\dagger}_{i2\sigma'}f_{i1\sigma'}f_{i2\sigma}  \nonumber \\
&+&g_{1}\sum_{i}( b^{\dag}_{i1}+b_{i1}) \hat{\tau}_{ix} 
+g_{2}\sum_{i}( b^{\dag}_{i2}+b_{i2})\hat{\tau}_{iz}   \nonumber \\
&+&\sum_{i\nu}\omega_{0\nu}b^{\dag}_{i\nu}b_{i\nu} 
\label{model}
\end{eqnarray}
with
\begin{eqnarray}
\hat{\tau}_{ix}&=&\sum_{\sigma}(f^{\dag}_{i1\sigma}f_{i2\sigma}+f^{\dag}_{i2\sigma}f_{i1\sigma}) \label{taux} \\
\hat{\tau}_{iz}&=&\sum_{\sigma}(f^{\dag}_{i1\sigma}f_{i1\sigma}-f^{\dag}_{i2\sigma}f_{i2\sigma}), \label{tauz}
\end{eqnarray}
where $c_{il\sigma}^{\dag}$ ($f_{il\sigma}^{\dag}$) is a creation operator for a conduction (c-)electron ($f$-electron) with orbital $l(=1,2)$ and spin $\sigma(=\uparrow ,\downarrow)$ at site $i$, and $\hat{n}_{fil\sigma}=f_{il\sigma}^{\dag}f_{il\sigma}$.
$b_{i\nu}^{\dag}$  is a creation operator for a Jahn-Teller phonon with mode $\nu(=1,2)$ at site $i$, where the normal coordinates for the Jahn-Teller phonons are given by $\hat Q_{i\nu}=1/\sqrt{2\omega_{0\nu}}(b_{i\nu}+b^{\dag}_{i\nu})$.
$\epsilon_{{\bf k}}$, $\epsilon_{f}$, and $V$ are the dispersion of c-electron, the atomic level of $f$-electron, and the c-$f$ hybridization, respectively. 
In the model eq.(\ref{model}), the Jahn-Teller phonons with the frequency $\omega_{0\nu}$ couple with the local orbital fluctuations of $f$-electrons, $\hat{\tau}_{ix}$ and $\hat{\tau}_{iz}$, via the electron-phonon coupling $g_{\nu}$ \cite{hotta2}. 
The model eq.(\ref{model}) also includes  the Coulomb interaction between $f$-electrons: the intra and inter orbital direct Coulomb $U$ and $U'$ and the exchange coupling $J$. 
For simplicity, we assume $\omega_{01}=\omega_{02}(\equiv\omega_{0})$, $g_{1}=g_{2}(\equiv g)$, $U=U'$ and $J=0$ in the present paper. 

To solve the model eq.(\ref{model}), we use the DMFT \cite{georges} in which the model is mapped onto an effective single impurity two-orbital Anderson model coupled with Jahn-Teller phonons \cite{koller}. 
The local Green's function $G_{fl\sigma}(\ii\omega_{n})$ and the local self-energy $\Sigma_{l\sigma}(\ii\omega_{n})$ for the $f$-electron satisfy the following self-consistency conditions: 
\begin{eqnarray*}
G_{fl\sigma}(\ii\omega_{n}) &=& \int d\epsilon\frac{\rho(\epsilon)}{\ii\omega_{n}-\epsilon_{f}-\Sigma_{l\sigma}(\ii\omega_{n})-\frac{V^2}{\ii\omega_{n}-\epsilon}} 
    \nonumber \\
   &=& \lbrack\tilde{G}_{fl\sigma}(\ii\omega_{n})^{-1}-\Sigma_{l\sigma}(\ii\omega_{n})\rbrack^{-1},
\end{eqnarray*}
where $\rho(\epsilon)$ is the density of states (DOS) for the $c$-electron, 
$\rho(\epsilon)=\sum_{\bf k}\delta(\epsilon-\epsilon_{{\bf k}})$. 
In the above equation, $\tilde{G}_{fl}(\ii\omega_{n})$ is the bare Green's function for the effective impurity Anderson model with $U=g=0$ in an effective medium which will be determined self-consistently. 
The effective impurity Anderson model with finite $U$ and/or $g$ is solved by using the ED method for a finite-size cluster to obtain $G_{fl\sigma}(\ii\omega_{n})$ and $\Sigma_{l\sigma}(\ii\omega_{n})$ at $T=0$ \cite{mitsumoto11,mitsumoto12,mitsumoto13,mitsumoto2,georges,koller,caffarel,ono}. 
In the present study, we use 4 site cluster and the cutoff of phonon number is set to be 9  for each Jahn-Teller mode. 
We assume a semielliptic DOS with the bandwidth $W=1$, $\rho(\epsilon)=2\sqrt{1-\epsilon^2}/\pi$, and we set $V=0.1$ and $\omega_{0}=0.01$. 
We concentrate our attention on the nonmagnetic state with $\Sigma_{l\sigma}(\ii\omega_{n})=\Sigma(\ii\omega_{n})$ at half-filling with $n_f=2$ and quarter-filling with $n_f=1$, where $n_f \equiv \sum_{l\sigma}\langle \hat{n}_{fil\sigma} \rangle$ is the average number of $f$-electrons per atom.

\section{Results}

\subsection{Renormalization factor}
\begin{figure}[tb]
\begin{center}
\includegraphics[angle=0,width=0.40\textwidth]{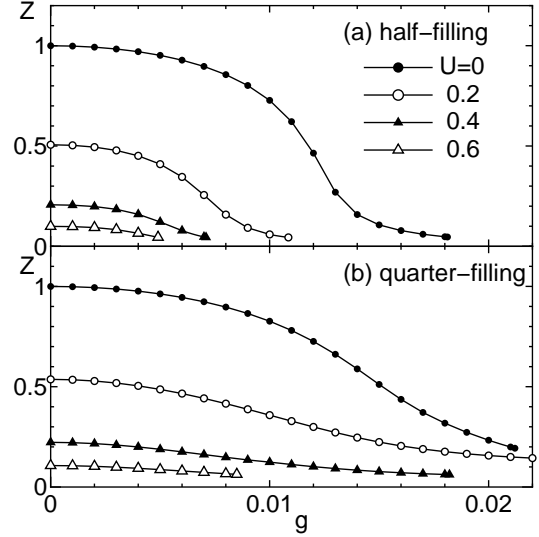}
\end{center}
\caption{
The renormalization factor $Z$ 
as a function of electron-phonon coupling constant $g$ for several values of the Coulomb interaction $U$ at half-filling (a) and quarter-filling (b).
}
\label{z}
\end{figure}

In Fig.\ref{z}, we plot the renormalization factor $Z=(1-\frac{\rm{d} \Sigma(\omega)}{\rm{d} \omega}|_{_{\omega=0}})^{-1}$  as a function of $g$ for several values of $U$ at half- and quarter-filling. 
When $g=0$, the effective mass of the quasiparticle $m^{*}/m=Z^{-1}$ increases with increasing $U$ and then the heavy fermion state with $m^{*}/m\gg 1$ is realized in the strong correlation regime. 
When the electron-phonon coupling increases, the effective mass increases for all values of $U$ resulting in the heavy fermion state due to the cooperative effect of the Coulomb interaction and the electron-phonon coupling.
This is a striking contrast to the case with the single-orbital periodic Anderson-Holstein model where the both effects are compete with each other as mentioned before \cite{mitsumoto11, mitsumoto12, mitsumoto13}. 
The effective mass $m^{*}/m$ for half-filling is considerably larger than that for quarter-filling as shown in Figs.\ref{z} (a) and (b). 
This is due to the effect of the local orbital fluctuation which is largely enhanced for half-filling as compared to quarter-filling as mentioned later.

\subsection{Local orbital fluctuation}
\begin{figure}[tb]
\begin{center}
\includegraphics[angle=0,width=0.42\textwidth]{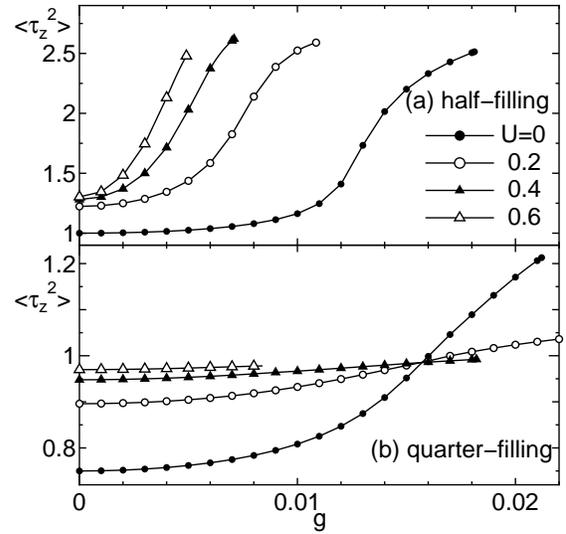}
\end{center}
\caption{
The local orbital fluctuation 
$\langle\tau_{z}^2\rangle$ 
as a function of $g$ for several values of $U$ at half-filling (a) and quarter-filling (b).
}
\label{orfl}
\end{figure}

Fig.\ref{orfl} shows the local orbital fluctuation written by
\begin{eqnarray}
\langle\tau_{z}^{2}\rangle=\langle\hat{\tau}_{iz}^{2}\rangle=\langle(\hat{n}_{fi1}-\hat{n}_{fi2})^2\rangle
\label{tauu}
\end{eqnarray}
with $\hat{n}_{fil}=\sum_{\sigma}\hat{n}_{fil\sigma}$
as a function of $g$ for several values of $U$, 
where 
$\langle\tau_{z}\rangle = \langle\tau_{x}\rangle=0$ 
in the absence of orbital order and 
$\langle\tau_{z}^{2}\rangle=\langle\tau_{x}^{2}\rangle$
due to the symmetry of $\mathrm{e_{u}}$ representation.
When $g=0$, $\langle\tau_{z}^{2}\rangle$ increases with increasing $U$ as previously obtained in the two-orbital periodic Anderson model \cite{sato}. 
When $g$ increases for $U\ne 0$, $\langle\tau_{z}^{2}\rangle$ steeply increases at half-filling, while it gradually increases at quarter-filling except for $U=0$ as shown in Figs. 2 (a) and (b). 
Remarkably, the increase in the local orbital fluctuation in the strong coupling regime is largely enhanced due to the Coulomb interaction for half-filling 
in contrast to the case for quarter-filling where it is suppressed. 
The difference between half-filling and quarter-filling in the local orbital fluctuation is due to that in the double-occupancy probabilities as shown below.

\subsection{Double occupancy probabilities}
\begin{figure}[tb]
\begin{center}
\includegraphics[angle=0,width=0.425\textwidth]{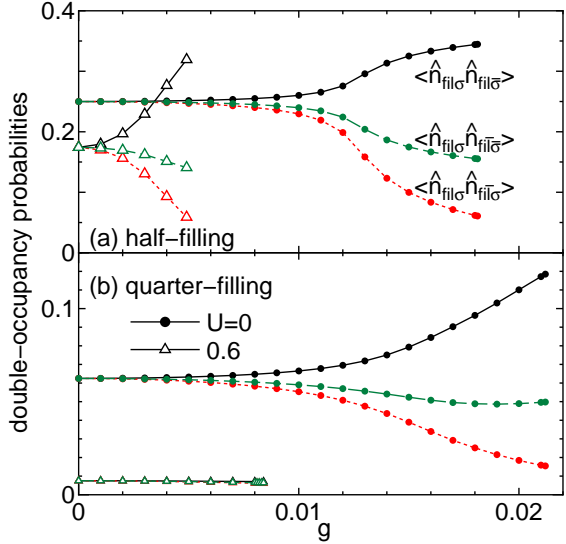}
\end{center}
\caption{
(Color online) 
The double-occupancy probabilities, 
$\langle\hat{n}_{fil\sigma}\hat{n}_{fil\bar{\sigma}}\rangle$ (solid lines), 
$\langle\hat{n}_{fil\sigma}\hat{n}_{fi\bar{l}\bar{\sigma}}\rangle$ (dashed lines) and $\langle\hat{n}_{fil\sigma}\hat{n}_{fi\bar{l}\sigma}\rangle$ (dotted lines), 
as functions of $g$ for $U=0$ and $0.6$ 
at half-filling (a) and quarter-filling (b), respectively.
}
\label{average}
\end{figure}

The local orbital fluctuation 
$\langle\tau_{z}^{2}\rangle$ 
given in eq.(\ref{tauu}) can be rewritten by
\begin{eqnarray}
\langle\tau_{z}^{2}\rangle=
\langle\hat{n}_{fi}\rangle
+2\sum_{l}\langle\hat{n}_{fil\uparrow}\hat{n}_{fil\downarrow}\rangle
-2\sum_{\sigma\sigma'}\langle\hat{n}_{fi1\sigma}\hat{n}_{fi2\sigma'}\rangle
\label{orbital}
\end{eqnarray}
with $\hat{n}_{fi}=\sum_{l}\hat{n}_{fil}$. 
To discuss the local orbital fluctuation in detail, we also plot the double-occupancy probabilities as functions of $g$ for several values of $U$ at half- and quarter-filling in Fig.\ref{average}, where 
$\langle\hat{n}_{fil\sigma}\hat{n}_{fil\bar{\sigma}}\rangle$, 
$\langle\hat{n}_{fil\sigma}\hat{n}_{fi\bar{l}\bar{\sigma}}\rangle$ and
$\langle\hat{n}_{fil\sigma}\hat{n}_{fi\bar{l}\sigma}\rangle$ 
are on-site correlation functions for intra-orbital spin-singlet, inter-orbital spin-singlet and inter-orbital spin-triplet, respectively. 
When $g=0$, the double-occupancy probabilities coincide to each other due to the spin-orbital symmetry with $U=U'$, and then, they are suppressed due to the electron correlation effect, especially in the case with quarter-filling as shown in Fig.\ref{average} (b). 
On the other hand, when $g\ne 0$, the degeneracy is resolved, and then, the intra-orbital spin-singlet correlation 
$\langle\hat{n}_{fil\sigma}\hat{n}_{fil\bar{\sigma}}\rangle$ 
is enhanced, while the inter-orbital spin-triplet correlation 
$\langle\hat{n}_{fil\sigma}\hat{n}_{fi\bar{l}\sigma}\rangle$ and 
the inter-orbital spin-singlet correlation 
$\langle\hat{n}_{fil\sigma}\hat{n}_{fi\bar{l}\bar{\sigma}}\rangle$ 
are suppressed due to the effect of $g$.
As the results, $\langle\tau_{z}^{2}\rangle$ given in eq.(\ref{orbital}) is largely enhanced for half-filling with large values of the double-occupancy probabilities, while it is slightly enhanced for quarter-filling with small values of those as shown in Fig.\ref{orfl}.

\subsection{Local charge fluctuation}
\begin{figure}[tb]
\begin{center}
\includegraphics[angle=0,width=0.40\textwidth]{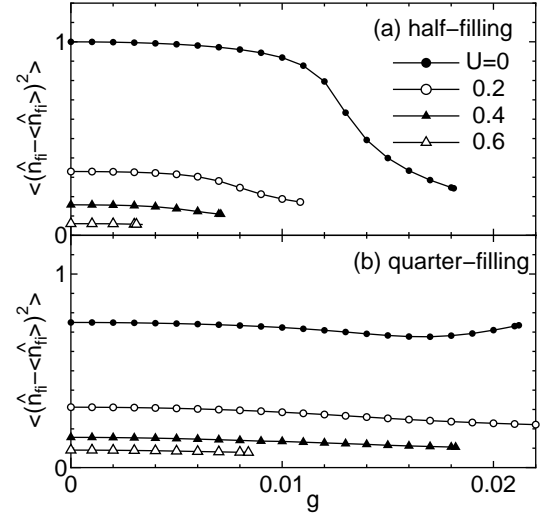}
\end{center}
\caption{
The valence fluctuation 
$\langle(\hat{n}_{fi}-\langle\hat{n}_{fi}\rangle)^{2}\rangle$
as a function of $g$ for several values of $U$
at half-filling (a) and quarter-filling (b).
}
\label{valence}
\end{figure}

The local charge fluctuation of $f$-electrons, i. e., the valence fluctuation is written by 
\begin{eqnarray}
\langle(\hat{n}_{fi}-\langle\hat{n}_{fi}\rangle)^{2}\rangle
&=&\langle\hat{n}_{fi}\rangle-\langle\hat{n}_{fi}\rangle^{2}
+2\sum_{l}\langle\hat{n}_{fil\uparrow}\hat{n}_{fil\downarrow}\rangle
\nonumber\\
&+&2\sum_{\sigma\sigma'}\langle\hat{n}_{fi1\sigma}\hat{n}_{fi2\sigma'}\rangle, 
\label{charge}
\end{eqnarray}
and is plotted as a function of $g$ for several values of $U$ at half- and quarter-filling in Fig.\ref{valence}.
When $g=0$, the  valence fluctuation decreases with increasing $U$ together with decreasing double-occupancy probabilities for both half- and quarter-filling due to the electron correlation effect. 
When $g\ne 0$, the valence fluctuation is suppressed or almost constant due to the effect of $g$ in contrast to the case with the single-orbital periodic Anderson model where the valence fluctuation is coupled with local phonons and is enhanced due to the effect of $g$\cite{mitsumoto11, mitsumoto12, mitsumoto13}. 
We note that the sign of the final term (the inter-orbital correlation) in eq.(\ref{charge}) is opposite to that in eq.(\ref{orbital}), and then, the changes in the intra- and inter- orbital correlations cancel with each other (see Fig.\ref{average}) resulting in the decrease in the valence fluctuation with $g$ in contrast to the increase in the local orbital fluctuation.

\subsection{Local moment}
\begin{figure}[tb]
\begin{center}
\includegraphics[angle=0,width=0.40\textwidth]{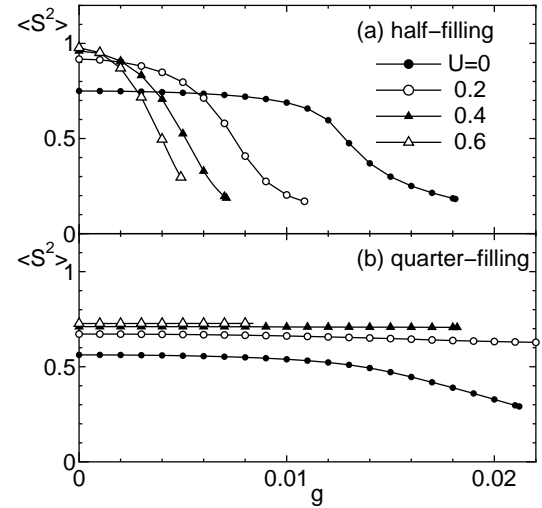}
\end{center}
\caption{
The local moment 
$\langle\mathbf{S}^{2}\rangle$ 
as a function of electron-phonon coupling constant $g$ for several values of $U$
at half-filling (a) and quarter-filling (b).
}
\label{spin}
\end{figure}

The local moment, 
$\langle\mathbf{S}^2\rangle=\langle\hat{\mathbf{S}}_{i}^2\rangle=\langle\hat{S}_{ix}^{2}+\hat{S}_{iy}^{2}+\hat{S}_{iz}^{2}\rangle$, 
written by
\begin{eqnarray*}
\langle\mathbf{S}^{2}\rangle
=\frac{3}{4}\left(\langle\hat{n}_{fi}\rangle+2\sum_{\sigma}\langle\hat{n}_{fi1\sigma}\hat{n}_{fi2\sigma}\rangle-2\sum_{ll'}\langle\hat{n}_{fil\uparrow}\hat{n}_{fil'\downarrow}\rangle\right)
\end{eqnarray*}
is plotted as a function of $g$ for several values of $U$ at half- and quarter-filling in Fig.\ref{spin}. 
When $g=0$, $\langle\mathbf{S}^2\rangle$ is enhanced due to the effect of $U$.
When $g$ increases, $\langle\mathbf{S}^2\rangle$ decreases with decreasing (increasing) spin-triplet (singlet) correlation as shown in Fig.\ref{average}. 
The decrease in $\langle\mathbf{S}^2\rangle$ with $g$ for half-filling is larger than that for quarter-filling, as the double-occupancy probabilities for half-filling are larger than those for quarter-filling (see Fig.\ref{average}).

\subsection{Local lattice fluctuation}
\begin{figure}[tb]
\begin{center}
\includegraphics[angle=0,width=0.445\textwidth]{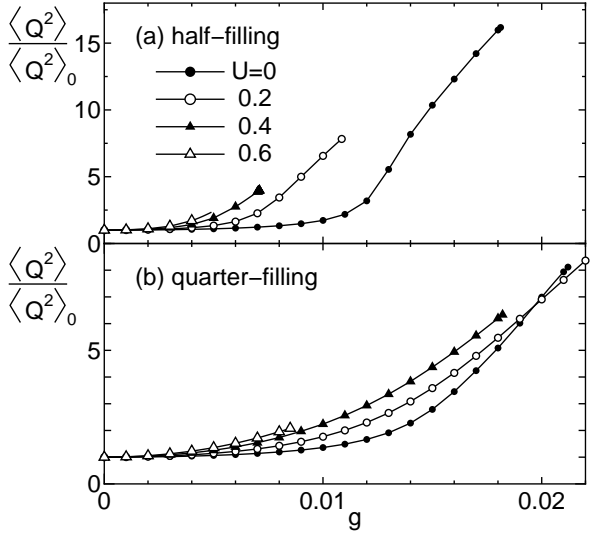}
\end{center}
\caption{
The normalized local lattice fluctuation $\langle Q^{2}\rangle /\langle Q^{2}\rangle_{0}$,
as a function of $g$ for several values of $U$
at half-filling (a) and quarter-filling (b).
}
\label{bx}
\end{figure}

The local lattice fluctuation is defined by 
$\langle Q^{2}_{\nu}\rangle =\langle  (\hat Q_{i\nu}-\langle \hat Q_{i\nu}\rangle)^{2}\rangle$ and $\langle Q^{2}_{1}\rangle=\langle Q^{2}_{2}\rangle(\equiv\langle Q^{2}\rangle)$ due to the symmetry of $\mathrm{E_{g}}$ modes. 
In Fig.\ref{bx}, we plot the normalized local lattice fluctuation, $\langle Q^{2}\rangle / \langle Q^{2}\rangle_{0}$, where $\langle Q^{2}\rangle_{0}=1/2\omega_{0}$ is the value for the zero-point oscillation with $g=0$. 
When $g$ increases, $\langle Q^{2}\rangle$ monotonically increases for all values of $U$. 
Remarkably, the increase in $\langle Q^{2}\rangle$ observed in the strong coupling regime is largely enhanced due to the Coulomb interaction especially for half-filling, where the orbital fluctuation coupled with the local phonons is largely enhanced. This is a striking contrast to the case with the single-orbital periodic Anderson model where $\langle Q^{2}\rangle$ is suppressed due to the Coulomb interaction as well as the valence fluctuation as mentioned before\cite{mitsumoto11, mitsumoto12, mitsumoto13}.

\subsection{$F$-electron spectral function}
\begin{figure}[tb]
\begin{center}
\includegraphics[angle=0,width=0.40\textwidth]{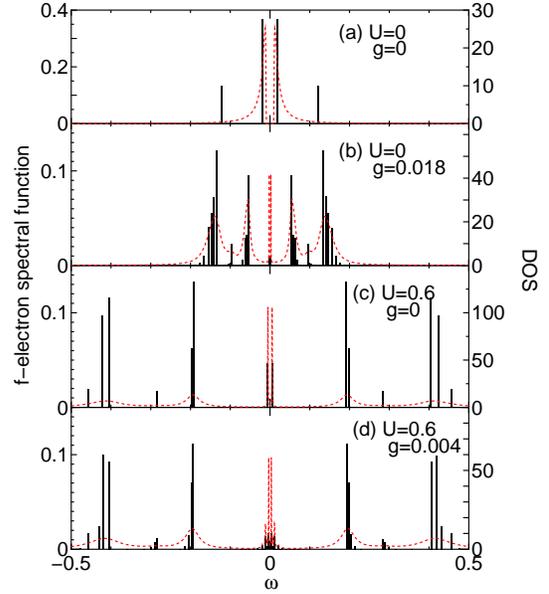}
\end{center}
\caption{
(Color online) 
The f-electron spectral function (vertical bars) and the DOS (dotted lines) as functions of $\omega$ for several values of $U$ and $g$ at half-filling, where the Fermi level is set to zero. 
}
\label{dosh}
\end{figure}

Fig.\ref{dosh} shows the $f$-electron spectral function as a function of $\omega$ at half-filling, where we also plot the analytic DOS for $U=g=0$ in Fig.\ref{dosh} (a) and the DOS evaluated by a Lorentzian broadening with width of $0.1|\omega|$ for finite $U$ and/or $g$ in Figs.\ref{dosh} (b)-(d). 
For $U=g=0$, the Fermi level is in the $c$-$f$ hybridization gap $\Delta_{cf}$ as shown in Fig.\ref{dosh} (a). 
For $U=0.6$ and $g=0$ shown in Fig.\ref{dosh} (c), the $c$-$f$ hybridization gap is renormalized as $\sim Z \Delta_{cf}$ with the renormalization factor $Z\sim0.099$ due to the strong electron correlation effect, i. e., the system becomes Kondo insulator. In this case,  we also observe the lower and upper Hubbard bands around $\omega = \pm U/2$. 

For $U=0$ and $g=0.018$ shown in Fig.\ref{dosh} (b), the $c$-$f$ hybridization gap is renormalized as $\sim Z \Delta_{cf}$ with $Z\sim0.048$ due to the effect of strong electron-phonon coupling. We note that, in the case with large phonon frequency, $\omega_{0}\gg W$, the system can be described by the effective negative-$U_{\mathrm{eff}}$ Anderson model with $|U_{\mathrm{eff}}|=2g^{2}/\omega_{0}$, 
where the lower and upper bands are formed around $\pm |U_{\mathrm{eff}}|/2$. 
In the present case with small $\omega_{0}\ll W$, however, the clear lower and upper Hubbard bands are not observed, instead incoherent sub-bands are observed together with the quasiparticle bands. 
For $U=0.6$ and $g=0.004$ shown in Fig.\ref{dosh} (d), the $c$-$f$ hybridization gap becomes narrower ($\sim Z \Delta_{cf}$ with $Z\sim0.065$) than that for $U=0.6$ and $g=0$ (see Fig.\ref{dosh} (c)) due to the cooperative effect of the Coulomb interaction and the electron-phonon coupling. In this case, the high energy structure of the lower and upper Hubbard bands observed for $U=0.6$ and $g=0$ is almost unchanged by the effect of $g$.

\begin{figure}[tb]
\begin{center}
\includegraphics[angle=0,width=0.40\textwidth]{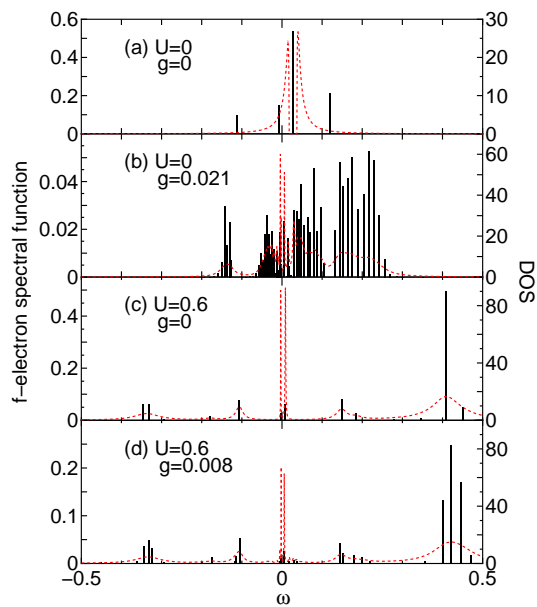}
\end{center}
\caption{
(Color online) 
The f-electron spectral function (vertical bars) and the DOS (dotted lines) as functions of $\omega$ for several values of $U$ and $g$ at quarter-filling, where the Fermi level is set to zero. 
}
\label{dosq}
\end{figure}

In Fig.\ref{dosq}, we plot the $f$-electron spectral function as a function of $\omega$ at quarter half-filling for $U=g=0$ ($Z=1$), $U=0$ and $g=0.021$ ($Z\sim0.20$), $U=0.6$ and $g=0$ ($Z\sim0.11$), and $U=0.6$ and $g=0.008$ ($Z\sim0.066$). 
The specific features of the $f$-electron spectral function and the DOS for quarter-filling shown in Fig.\ref{dosq} are almost the same as those for half-filling shown in Fig.\ref{dosh} except that the DOS is asymmetric in contrast to the case with half-filling where the DOS is symmetric. Then, the system becomes heavy fermion metal with large effective mass $m^*/m=Z^{-1}$, instead of the Kondo insulator with narrow $c$-$f$ hybridization gap $\sim Z \Delta_{cf}$ realized for half-filling.

\subsection{Phonon spectral function}
\begin{figure}[tb]
\begin{center}
\includegraphics[angle=0,width=0.45\textwidth]{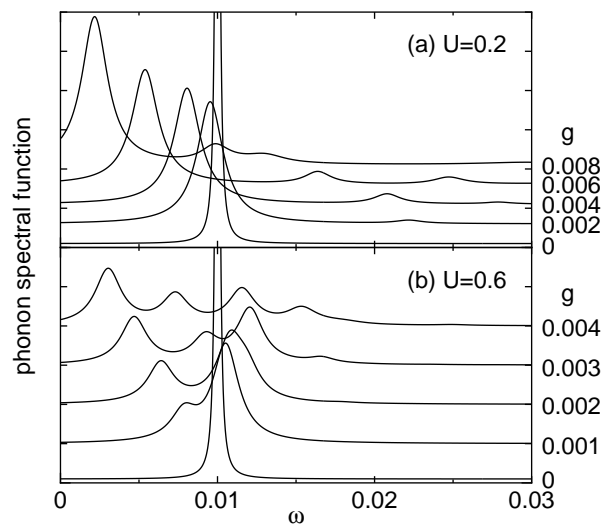}
\end{center}
\caption{
The phonon spectral function as a function of $\omega$ at half-filling for several values of $g$ with $U=0.2$ (a) and $U=0.6$ (b).
}
\label{phonon1_h}
\end{figure}

\begin{figure}[tb]
\begin{center}
\includegraphics[angle=0,width=0.45\textwidth]{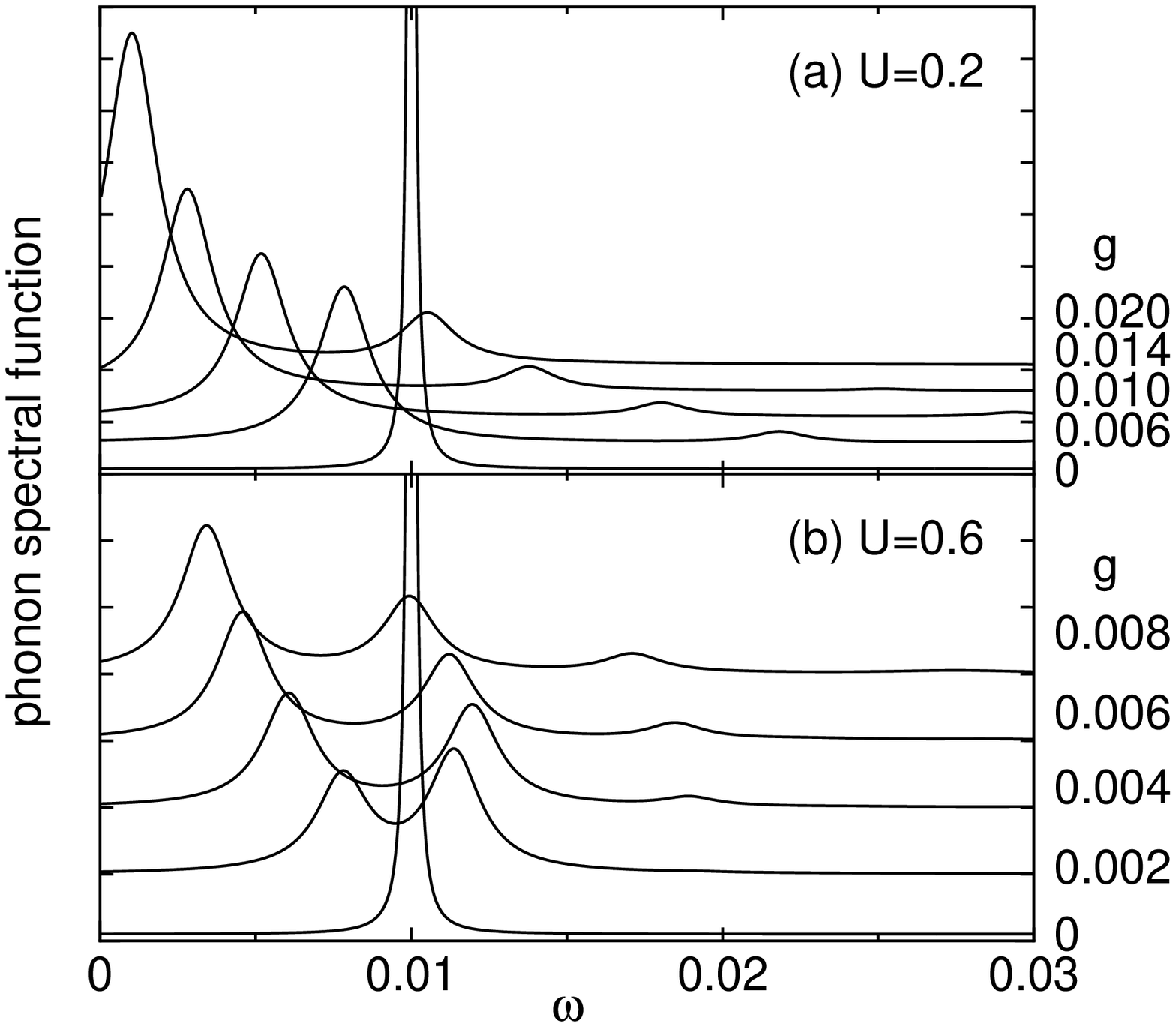}
\end{center}
\caption{
The phonon spectral function as a function of $\omega$ at quarter-filling for several values of $g$ with $U=0.2$ (a) and $U=0.6$ (b).
}
\label{phonon1_q}
\end{figure}

Finally, we calculate the local phonon Green's function defined by 
$D_{\nu}(\tau)=-\langle T_{\tau}b_{i\nu}(\tau)b_{i\nu}^{+}(0)\rangle$, 
whose spectral function is plotted as a function of $\omega$ for half-filling in Fig.\ref{phonon1_h} and for quarter-filling in Fig.\ref{phonon1_q}, respectively. 
When $g$ increases, the lowest excited energy shifts to low energy as previously observed for the single-orbital periodic Anderson-Holstein model \cite{mitsumoto11, mitsumoto12, mitsumoto13}. 
For a small value of $U=0.2$, the lowest excited state has a dominant contribution to the spectral weight. 
On the other hand, for a large value of $U=0.6$, the spectral weight of the lowest excited state is relatively small and a considerable amount of the spectral weight remains around the original phonon level. 
This result is consistent with that obtained for the Hubbard-Holstein model in the strong coupling regime, where the sharp soft phonon mode with a large spectral weight is observed  for small-$U$, while the broad soft phonon mode with a small spectral weight is observed for large-$U$ \cite{koller}.

\begin{figure}[tb]
\begin{center}
\includegraphics[angle=0,width=0.42\textwidth]{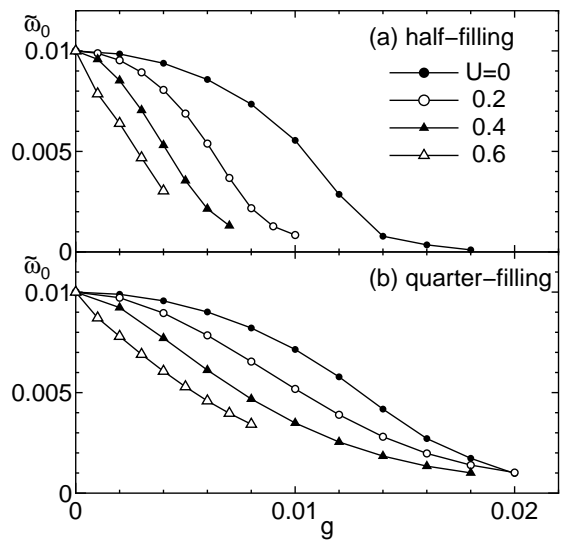}
\end{center}
\caption{
The lowest excited energy $\tilde{\omega}_{0}$ in the phonon spectral function as a function of $g$ for several values of $U$  at half-filling (a) and quarter-filling (b).
}
\label{phonon2}
\end{figure}

To see the energy shift in more detail, we show the lowest excited energy $\tilde{\omega}_{0}$ in the phonon spectral function as a function of $g$ in Fig.\ref{phonon2}.
Remarkably, the softening, i. e., the decrease in $\tilde{\omega}_{0}$, observed in the strong coupling regime is largely enhanced due to the Coulomb interaction especially for half-filling in contrast to the case with the single-orbital periodic Anderson-Holstein model where the softening is suppressed due to the Coulomb interaction\cite{mitsumoto11, mitsumoto12, mitsumoto13}.

\section{Summary and Discussions}
In Summary, we have investigated the two-orbital periodic Anderson model coupled with the two-fold degenerate Jahn-Teller phonon by using the dynamical mean-field theory, and have found that the heavy fermion state of nonmagnetic origin is realized due to the cooperative effect of the Coulomb interaction $U$ and the electron-phonon coupling $g$. 
The specific features of the heavy fermion state for large $U$ and $g$ are as follows: 
(1) The local orbital and lattice fluctuations are enhanced, while the local charge (valence) and spin fluctuations are suppressed. 
(2) The sharp soft phonon mode with a large spectral weight is observed for small $U$, while the broad soft phonon mode with a small spectral weight is observed for large $U$. 
(3) The cooperative effect for half-filling with $n_f=2$ is more pronounced than that for quarter-filling with $n_f=1$. 

In the absence of $g$, when $U$ increases, the effective mass increases with increasing spin and orbital fluctuations together with decreasing charge fluctuation. Then, the heavy fermion state of magnetic origin with large (small) spin and orbital (charge) fluctuations is realized for large $U$ due to the strong correlation effect. 
When we increase $g$ for a fixed value of $U$, the effective mass further increases with increasing orbital and lattice fluctuations together with decreasing spin and charge fluctuations. 
Then, the heavy fermion state of nonmagnetic origin with large (small) orbital and lattice (spin and charge) fluctuations is realized for large $U$ and $g$ due to the cooperative effect of the strong correlation and the strong coupling. 
In this heavy fermion state, the phonon spectral function shows the broad soft phonon mode while the $f$ electron spectral function is almost unchanged due to the effect of $g$ except further narrowing of the low energy quasiparticle bands. 

In the two-orbital periodic Anderson model with the coupling $g$ between the local orbital fluctuation and the Jahn-Teller phonon, the orbital fluctuation is enhanced due to the both effects of $U$ and $g$, and then the heavy-fermion state is realized due to the cooperative effect. 
This is a striking contrast to the case with the single-orbital periodic Anderson model with the coupling $g$ between the local charge fluctuation and the local phonon, where the spin (charge) fluctuation is enhanced (suppressed) due to the effect of $U$, while the charge (spin) fluctuation is enhanced (suppressed) due to the effect of $g$, and then the effects of $U$ and $g$ on the heavy-fermion state compete with each other \cite{mitsumoto11, mitsumoto12, mitsumoto13}. 
As the absolute value of the local orbital fluctuation for  $n_f=2$ is larger than that for $n_f=1$, the cooperative effect of $U$ and $g$ for $n_f=2$ is more pronounced than that for $n_f=1$. 

Although the present model is the simplest version, the multi orbital periodic Anderson model coupled with the Jahn-Teller phonons has the potential to account for the experimental observations. 
The heavy fermion state of nonmagnetic origin due to the cooperative effect of $U$ and $g$ seems to be consistent with the magnetically robust heavy fermion state observed in the filled skutterudite SmOs$_4$Sb$_{12}$. 
The enhancement of the orbital fluctuation and the anomalous phonon softening due to the cooperative effect of $U$ and $g$ seemes to be responsible for the extra softening of the low temperature elastic constant which is not reproduced only due to the quadrupolar susceptibility in the CEF state observed in PrOs$_4$Sb$_{12}$. 
To discuss these compounds in detail, however, we need to consider more realistic models including the effects such as the spin-orbit interaction, the Hund's rule coupling and the crystalline-electric-field. 
Within the j-j coupling scheme, Hotta has developed the realistic three orbital Anderson model with $\Gamma_{7}$ doublet and $\Gamma_{8}$ quartet CEF states simulating filled skutterudite compounds \cite{hotta2}. 
It is interesting to extend our approach to such realistic models and is now under the way.

\section*{Acknowledgment}

The authors thank T. Goto, Y. Nemoto and T. Hotta for many useful comments and discussions. 
This work is supported by a Grant-in-Aid for Scientific Research in a Priority Area "Skutterudite" (No. 15072202) from the Ministry of Education, Culture, Sports, Science and Technology.

\end{document}